\documentclass[12pt]{article}

\usepackage{authblk}
\usepackage{amsmath}
\usepackage{float}

\usepackage{graphicx}
\usepackage{dcolumn}
\usepackage{bm}

\usepackage[natbib=true,style=numeric,sorting=none]{biblatex}
\usepackage[right=2.5cm, bottom=2.5cm, left=2.5cm, top=2cm]{geometry}

\usepackage{xcolor}

\usepackage{lineno}

\usepackage{amssymb}

\begin{document}

\title{FPGA-based TDC for SiPM Timing and Amplitude Measurements}

\author[1]{\small Lucas Finazzi\footnote{Both authors contributed equally to this work. Corresponding author: lfinazzi@unsam.edu.ar}}
\author[1]{Felipe Soriano$^*$}
\author[1]{Leandro Gagliardi}
\author[1]{Federico Golmar}

\affil[1]{Instituto de Ciencias Físicas, UNSAM-CONICET, Buenos Aires, Argentina}

\date{\today}

\maketitle 

\begin{abstract}
    Silicon photomultipliers (SiPMs) are widely used in photon-counting applications, such as positron emission tomography or particle physics, where precise timestamps and photoelectron number information are both required. In this work, a Time-to-Digital Converter with amplitude measurement capabilities was designed using an Artix-7 XC7A35T FPGA. For amplitude measurements, the internal xADC of the FPGA was chosen over other alternatives for simplicity. This design has two channels and can operate in timestamp-only mode or timestamp + amplitude mode. In turn, channels can be operated independently or in coincidence. The Integral Non-Linearity and Differential Non-Linearity of the design were studied, along with its time resolution, which is $(19 \pm 1)$~ps. Using a CAEN DT5810 pulse emulator, the design was tested further in timing-only mode, timing + amplitude mode and in coincidence mode to showcase its performance when using Poisson distributed SiPM-like pulses of various amplitudes.
\end{abstract}

\section{Introduction} \label{sec:intro}

Silicon Photomultipliers (SiPMs) are optical sensors~\cite{sipm_review1, sipm_review2} that are capable of single photon detection. They are comprised of many avalanche photodiodes in parallel, which makes them photon number resolving. Due to these capabilities, these detectors are widely used in several scientific fields, including medicine~\cite{pet1, pet2}, particle physics~\cite{hep1, hep2, hep3}, quantum optics~\cite{qo1, finazzi_bunching_2024}, astrophysics~\cite{ap1, ap2}, space applications~\cite{gecam, finazzi_sipic_2026, finazzi_begonia_2024}, among others.

In many of these applications, the simultaneous measurement of timing and energy/amplitude is essential. Precise timing enables accurate event timestamps in Time-of-Flight (TOF) systems, and energy/amplitude measurements enable spectroscopy, particle identification, and event discrimination applications (for example in scintillation-based detectors~\cite{sipm_scint}). Achieving both measurements with high resolution poses significant challenges for readout electronics design.

Existing solutions typically rely on Application-Specific Integrated Circuits (ASICs), which provide excellent performance and high channel density~\cite{radioroc}. However, these devices lack flexibility and have significant development and integration costs. FPGA-based readout systems have been increasingly explored as an alternative for SiPM signal acquisition, enabling the implementation of flexible and reconfigurable architectures~\cite{fpga_sipm}.

This work presents an FPGA-based readout platform based on Xilinx Artix-7 XC7A35T for the simultaneous acquisition of timing and amplitude information of SiPM events. Timing information in an event is obtained with an FPGA-based Time-to-Digital Converter (TDC) with a Tapped Delay Line (TDL) architecture, while the amplitude information is obtained using the internal xADC included in Artix-7 parts. This approach seeks to reduce complexity and aims to make the design software defined, rather than hardware defined, which allows for easy prototyping with Artix-7 development boards.

\subsection{Time-to-Digital Converters based on Tapped Delay Lines}

\begin{figure}[H]
    \centering
    \includegraphics[width=0.85\textwidth]{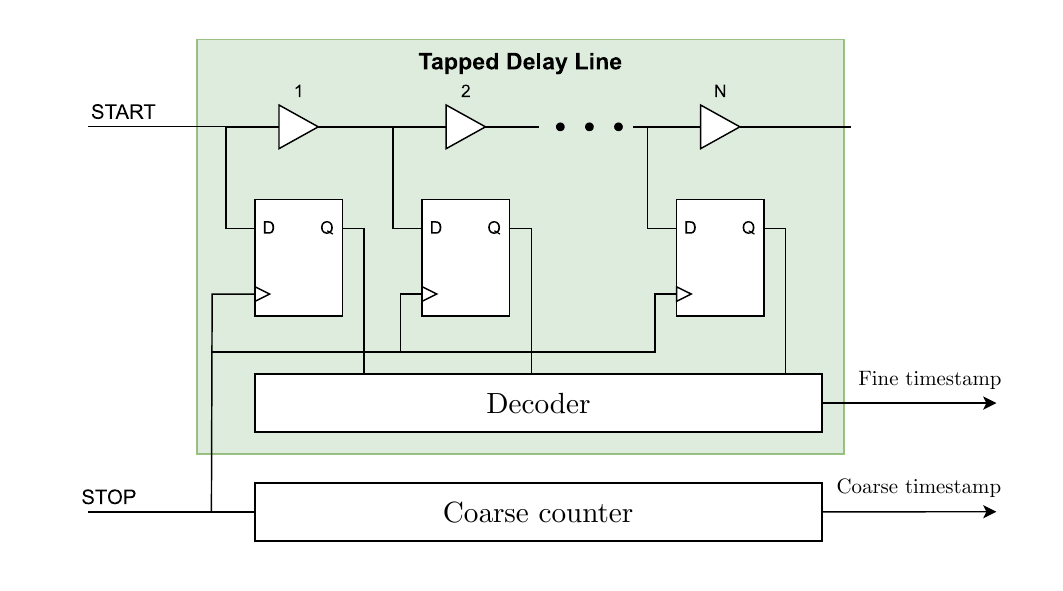}
    \caption{When an input signal occurs (START signal), it propagates through the taps, and each introduces a small, average propagation delay $\tau_p$. The next edge of the system clock acts as a STOP signal for a bank of flip-flops that simultaneously sample the state of every tap and give information about the number of total taps $n$ that the signal went through.}
    \label{fig:tdl}
\end{figure}

A Tapped Delay Line (TDL) is a structure used to delay an input signal. The signal propagates through of a series of stages called ``taps'', and it is possible to determine the corresponding time delay of this signal by measuring the number of taps traversed. This architecture allows the measurement of delays shorter than the system clock's period. A graphical depiction of a tapped delay line is shown in Figure~\ref{fig:tdl}.

When an input signal occurs (START signal), it propagates through the taps, and each contributes a small, average propagation delay $\tau_p$. The next edge of the system clock acts as a STOP signal for a bank of flip-flops that simultaneously latch the state of every tap and this gives information about the total number of taps $n$ that the signal went through. The number of total taps is chosen to cover a whole clock period. This way, the absolute time-tag for a START event can be calculated with a coarse counter (given by the system clock) and a fine counter (measured with the TDL). The time-tag for an event can be calculated as

\begin{equation}
    T_{\text{meas}} = \overbrace{\frac{N_{\text{clk}}}{f_{\text{clk}}}}^{\text{coarse time}} - \overbrace{n \cdot \tau_p}^{\text{fine time}}, \qquad \sigma_T^2 = \sigma_{\text{jitter}}^2 + \frac{\tau_p^2}{12} \ ,
\end{equation}

\noindent where $f_{clk}$ is the frequency of the system clock, $N_{\text{clk}}$ is the number of coarse clocks elapsed, $\sigma_T$ is the uncertainty of the measured time, and $\sigma_{\text{jitter}}$ is the jitter between START and STOP signals (which has a contribution from the clock jitter and SiPM signal jitter). In FPGA implementations, the delay elements are built from logic primitives whose propagation delays are small. In Xilinx Artix-7 devices, the \texttt{CARRY4} carry-chain cell is the standard choice for TDL implementations. These cells are laid out in fixed columns within the FPGA fabric, with direct fast connections between adjacent cells in the same column, ensuring consistent and predictable inter-tap delays.

In practice, the delay $w_k$ of each individual tap deviates from the nominal value $\tau_p$ due to manufacturing imperfections and operating conditions. The linearity of a TDL-TDC is therefore characterized by two figures of merit. The \textit{Differential Non-Linearity} (DNL) quantifies the k$^{\text{th}}$ tap deviation from the average tap width $\tau_p$, while the \textit{Integral Non-Linearity} (INL) captures the accumulated effect of these deviations up to the last measurement tap $n$. These are calculated as

\begin{equation}
    \text{DNL}[k] = \frac{w_k - \tau_p}{\tau_p} \ , \qquad \text{INL}[n] = \sum_{k=1}^n \text{DNL}[k] = \frac{\sum_{k=1}^{n} w_k - n \cdot \tau_p}{\tau_p} \ .
\end{equation}

While DNL reflects local imperfections (tap by tap), INL reflects how far the overall delay function drifts from an ideal straight line (each tap having the same delay). The main source of these non-linearities is sensitivity to process, voltage, and temperature (PVT) variations across the FPGA fabric. 

\subsubsection{Tap Calibration}

This PVT effect can be mitigated through an online tap calibration, where a correction lookup table is constructed, mapping each tap index to a calibrated time value and effectively flattening the delay function. For this purpose, an uncorrelated input source is used and a histogram of taps traversed is accumulated. Because the input events are uncorrelated with the sampling clock, their arrival phase is uniformly distributed over one clock period, so each tap is populated in proportion to its physical time width rather than its ideal width $\tau_p$. The resulting bin counts give an estimate of the delay of tap $k$, which is

\begin{equation}
    w_k = \frac{N[k]}{N_{\text{total}}} \times \frac{1}{f_{\text{clk}}} \ ,
    \label{eq:calib}
\end{equation}

\noindent where $N[k]$ is the number of events in bin $k$ and $N_{\text{total}}$ is the total number of events in this histogram. 

\section{FPGA Design} \label{sec:architecture}

The target device of this work is the Artix-7 XC7A35TICSG324-1L. The TDC core in this work builds upon (and adapts) the TDL-based TDC presented in a previous work~\cite{adamic}, which implements continuous timestamp acquisition on a Zynq-7010 device. For simultaneous amplitude measurements of incoming events, the internal xADC of the FPGA was used. The control logic was also changed and adapted to enable measurements with two channels in coincidence mode.

The Zynq-7010 is a system-on-chip that couples programmable logic to a hard ARM processor system, which was used in the original work for configuration and event readout. The Artix-7 offers no such processor, so design programmability and event readout were implemented in the FPGA fabric by embedding a MicroBlaze soft-core processor. In addition, the carry-chain placement constraints that define the delay line are specific to the device and to its floor plan, and had to be re-derived for the new part used in this work.

\subsection{Operation Modes}

The design in this work can operate in different modes. In all of them, the design records timestamps from one or both channels independently. The different operation modes differ in what is stored along each timestamp and in the condition that enables event saving during acquisition:

\begin{itemize}
    \item \textbf{CALIBRATION}: continuous acquisition used to build and store the bin-width calibration table of the TDL.
    \item \textbf{TDC}: continuous acquisition of timestamps only.
    \item \textbf{TAC (Time-and-Amplitude Converter)}: each timestamp is paired with an xADC measurement of the amplitude of the input signal.
    \item \textbf{COINCIDENCE}: acquisition is enabled only when both channels register an event within a user-defined coincidence window. Can be used with timestamp-only or timestamp + amplitude measurements.
\end{itemize}

\subsection{Architecture}

\begin{figure}[htbp]
    \centering
    \includegraphics[width=1.0\textwidth]{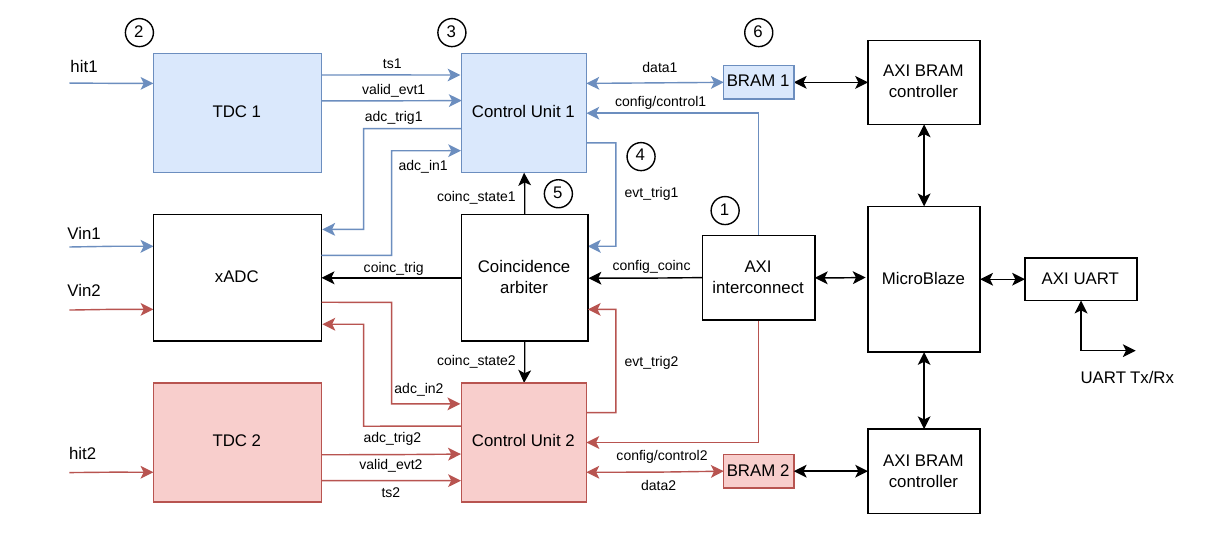}
    \caption{FPGA digital architecture. The design has two input channels, which can be operated individually or in coincidence (managed by the coincidence arbiter). Numbers showcase the flow of the system during an event in COINCIDENCE mode. The system is controlled through a MicroBlaze soft-core processor in the FPGA fabric.}
    \label{fig:arch}
\end{figure}

A block diagram of the FPGA digital architecture is shown in Figure~\ref{fig:arch}. The system clock used was 200~MHz to guarantee timing closure and 252 taps were used per channel to cover a complete clock period, corresponding to a theoretical bin width of 19.8~ps.

The design has two input channels (each channel has a digital input and an analog input), and each of them groups the blocks associated with a single SiPM detector input: a TDC, a dedicated control unit, and a BRAM block for intermediate data storage. Three blocks are shared between both channels: the coincidence arbiter, which enables acquisition when the channels are operated together; the internal xADC, which measures pulse amplitude; and the MicroBlaze soft-core processor, which is used to configure the design and read out events saved. The numbers in Figure~\ref{fig:arch} follow the flow of an event in COINCIDENCE mode:

\begin{enumerate}
    \item The user selects COINCIDENCE mode and sets the coincidence window length by writing to the control registers of both channels.
    \item The TDC 1 block receives an incoming hit signal.
    \item A timestamp is generated by the TDC and is transferred to Control Unit 1 for its temporal storage.
    \item The Control Unit 1 lets the Coincidence Arbiter know that an event was registered in channel 1 with the \texttt{evt\_trig1} signal.
    \item The arbiter opens a coincidence window and waits for an event trigger from channel 2 by monitoring the \texttt{evt\_trig2} signal. After the window is closed, an event in channel 2 was either registered or not and the arbiter notifies this to the control units with the \texttt{coinc\_state} signals.
    \item If there was a valid event in both channels during the coincidence window, data is saved into dedicated BRAM to be later retrieved by the MicroBlaze upon a user command. If not, the received event is ignored and not saved.
\end{enumerate}

The coincidence arbiter functionality is symmetric, which means that both channels can register a first event in COINCIDENCE mode. When amplitude measurements are enabled in this mode, the xADC is operated in event-driven mode at 1\,MS/s. Each time a timestamp is generated by a channel, its control unit requests a conversion after a user-defined delay, so that the sample is taken near the maximum of the analog input signal and its amplitude is registered together with the measured timestamp. The internal xADC was chosen for simplicity; a faster external ADC could be incorporated in future designs that require faster conversion times. The minimum size of the coincidence window that can be chosen is twice the clock period (10~ns).

\begin{figure}[htbp]
    \centering
    \includegraphics[width=1.0\textwidth]{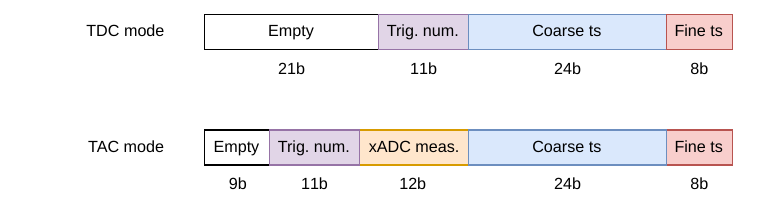}
    \caption{Frames saved in BRAM for each channel depending on operating mode. TDC mode just saves timing information, while TAC mode saves the xADC measurement along with the timestamp. COINCIDENCE mode can have either event structure, depending if amplitude measurements are enabled. The trigger number is also saved in all modes for later channel synchronization in case of a lost event.}
    \label{fig:frame}
\end{figure}

Depending on operation mode, each event saved in BRAM is composed of 64 bits and their structure is shown in Figure~\ref{fig:frame}. Each channel can store 2048 events before readout is required. Each BRAM block was created using Xilinx's \textit{Block Memory Generator} IP.

Communication with a PC is done via UART, with the MicroBlaze translating user commands into internal control signals and exposing the system state for monitoring and readout from the BRAMs is performed through an AXI BRAM Controller interface. In use cases where UART is a limiting factor in event readout speed, the design allows usage of a 100~Mbps Ethernet interface. Embedding a soft-core processor rather than implementing the communication logic purely in HDL was motivated by extensibility, with a view to future developments that may require on-board data processing or real-time veto logic. The application running on the MicroBlaze soft-core was programmed using Vitis. 

\subsection{Resource Utilization and Layout}

The complete system, comprised by the two TDC channels, the coincidence and acquisition logic, the storage BRAMs and the MicroBlaze subsystem was synthesized and implemented on the target device described above. Table~\ref{tab:fpga_utilization} summarizes the post-implementation resource utilization together with the total on-chip power estimated by Vivado.

\begingroup
\setlength{\belowcaptionskip}{10pt}
\begin{table}[H]
    \centering
    \caption{FPGA resource utilization and power consumption of the implemented design. The LUT count includes logic, distributed RAM and shift register usage. The power figure corresponds to an estimation after routing, and is 174~mW (dynamic) and 63~mW (static).}
    \label{tab:fpga_utilization}
    \begin{tabular}{lcc}
        \hline
        Resource & Units used & Percentage \\
        \hline
        LUTs & 2672 & 12.85\% \\
        Block RAM & 40 & 80.00\% \\
        I/O & 11 & 5.24\% \\
        Total on-chip power & 236~mW & --- \\
        \hline
    \end{tabular}
\end{table}
\endgroup

The design occupies a small fraction of the available logic (roughly 13\% of the LUTs) but uses 80\% of the Block RAM available. This is the main resource constraint for any future extension of the system. Only 8 of the 40 blocks in use correspond to the acquisition buffers (four per channel). The remaining 32 are the local memory of the MicroBlaze subsystem. Adding further channels would consequently not demand a proportional increase in memory, as the channel data buffers themselves are inexpensive, and the more direct way to recover Block RAM is to reduce the local memory allocated to the soft-core processor. If that were not possible, each channel would have to be given a shorter buffer, and data would need to be read out more frequently.

\begin{figure}[htbp]
    \centering
    \includegraphics[width=0.25\textwidth, angle=270]{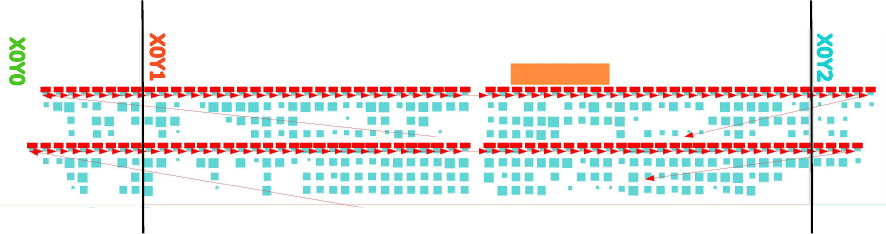}
    \caption{Example of tap placement for TDL of both channels. Placement constraints were used in Vivado to guarantee placement of taps in a straight line side by side to maximize timing uniformity.}
    \label{fig:tap_layout}
\end{figure}

Beyond the reported utilization, the physical placement of the delay-line elements is critical to the performance of the TDC, and was therefore constrained manually. The taps in the TDLs used were placed side by side in a straight line, with controlled spacing between them to maximize timing uniformity. An example of this placement in the design is shown in Figure~\ref{fig:tap_layout}.

\subsection{Tap calibration}

\begin{figure}[H]
    \centering
    \includegraphics[width=0.65\textwidth]{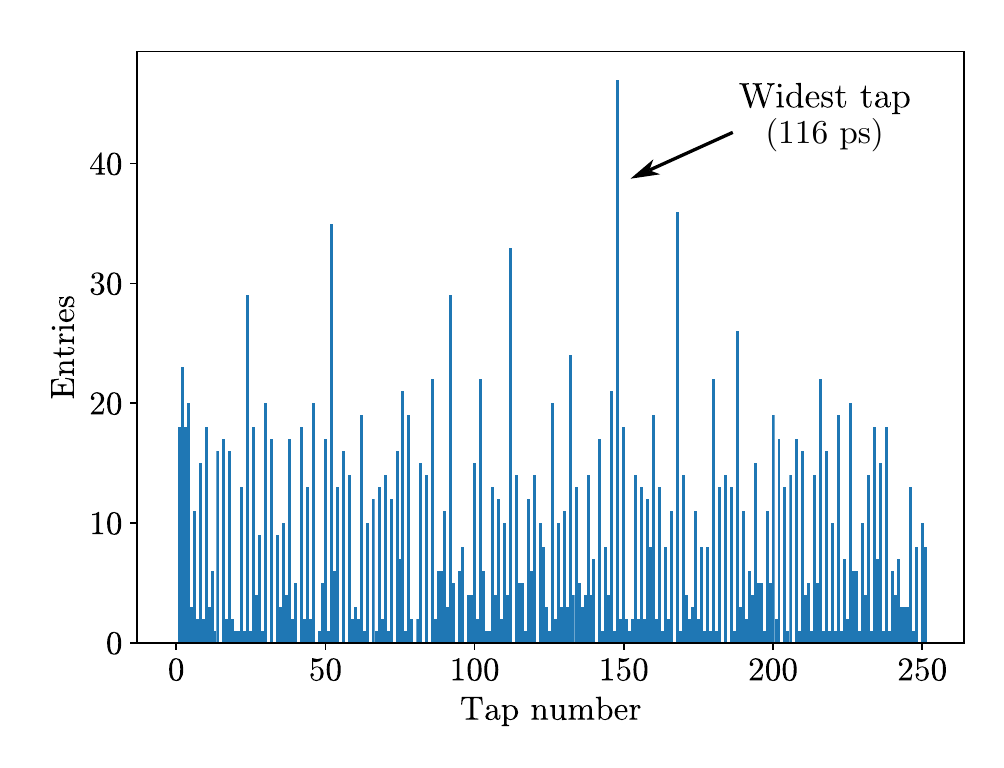}
    \caption{Histogram of fine timestamps measured for an ONSemi MicroFC-10035 SiPM (in dark conditions, at 5~V overvoltage and at 24~$^{\text{o}}$C) with channel 1. Wide bins with more delay than the average result in additional events in that given bin. The delay of each tap is calculated with equation~(\ref{eq:calib}) and saved in MicroBlaze dedicated memory. The widest tap in the design (for channel 1) was 116~ps.}
    \label{fig:calib_taps}
\end{figure}

As previously stated, an online calibration is needed to mitigate PVT variations in TDL-based TDCs. To do this, each channel must be fed an uncorrelated pulsed source (an SiPM signal, for example, which follows a random Poisson process) and a histogram of the fine timestamps recorded must be built. The CALIBRATION mode is used for this purpose before any measurements are performed. An example of a histogram built from the calibration measurements is shown in Figure~\ref{fig:calib_taps}.

\section{Experimental Setup} \label{sec:exp_setup}

An Arty-A7-35T development board was used for implementation of the design and experimental validation tests. A pMOD to SMA adapter was designed and used for all input signals. These input signals were generated with a CAEN DT5810 pulse emulator. This emulator allows the generation of pulses with user-controlled energy/amplitude profiles, timing statistics, noise, among other things. Using this instrument allows for a more exhaustive characterization of the FPGA design under real input conditions. This emulator has two independent channels, and each of them was used for a different channel of the FPGA design previously described. The emulator's analog output was connected directly to the analog inputs of the design (Vin1, Vin2) and a TLV3501 comparator was used to convert these pulse signals to LVCMOS 3.3~V for the digital TDC inputs (hit1, hit2). This comparator was chosen for its low propagation delay (4.5~ns) and fast rise time (1.5~ns), which helps preserve timing precision.

The following experiments were performed with this configuration (except the time resolution measurements, in which the same single channel of the pulse emulator was connected to both FPGA input channels):

\begin{itemize}
    \item CALIBRATION measurements for both channels, including the construction of the calibration table and INL/DNL measurements. 
    \item Measurement of the time resolution of the system. As the same physical signal was connected to both FPGA channels, any timing jitter is caused by the FPGA design itself and determines its time resolution.
    \item TDC mode measurements with a single channel to compare the measured and expected time distribution of the input signal.
    \item TAC mode measurements with a single channel to assess the amplitude measurement along with the time-tag for each event.
    \item COINCIDENCE mode measurements to test the performance of the user-defined coincidence window and the rejection of events outside that time interval.
\end{itemize}

In all cases, SiPM pulses from the emulator had the characteristics presented in Table~\ref{tab:afe_comparison} and a real finger spectrum was used to emulate SiPM crosstalk and generate pulses of different amplitudes (or photoelectrons). The finger spectrum used was measured from an ONSemi MicroFC-10035 SiPM at 5~V overvoltage and at 24~$^{\text{o}}$C. Poisson statistics were chosen for the pulses, with a rate of 20~kHz.

\begin{table}[H]
    \setlength{\belowcaptionskip}{10pt}
    \centering
    \caption{Parameters used by pulse emulator for the generation of input signals.}
    \label{tab:afe_comparison}
    \begin{tabular}{lcc}
        \hline
        Parameter & Value \\
        \hline
        pulse FWHM & 2~$\mu$s \\
        1 p.e. amplitude & 500~mV \\
        Crosstalk probability & 16\% \\
        \hline
    \end{tabular}
\end{table}

\section{Results} \label{sec:results}

\subsection{Timing performance and calibration}

\begin{figure}[H]
    \centering
    \includegraphics[width=1.0\textwidth]{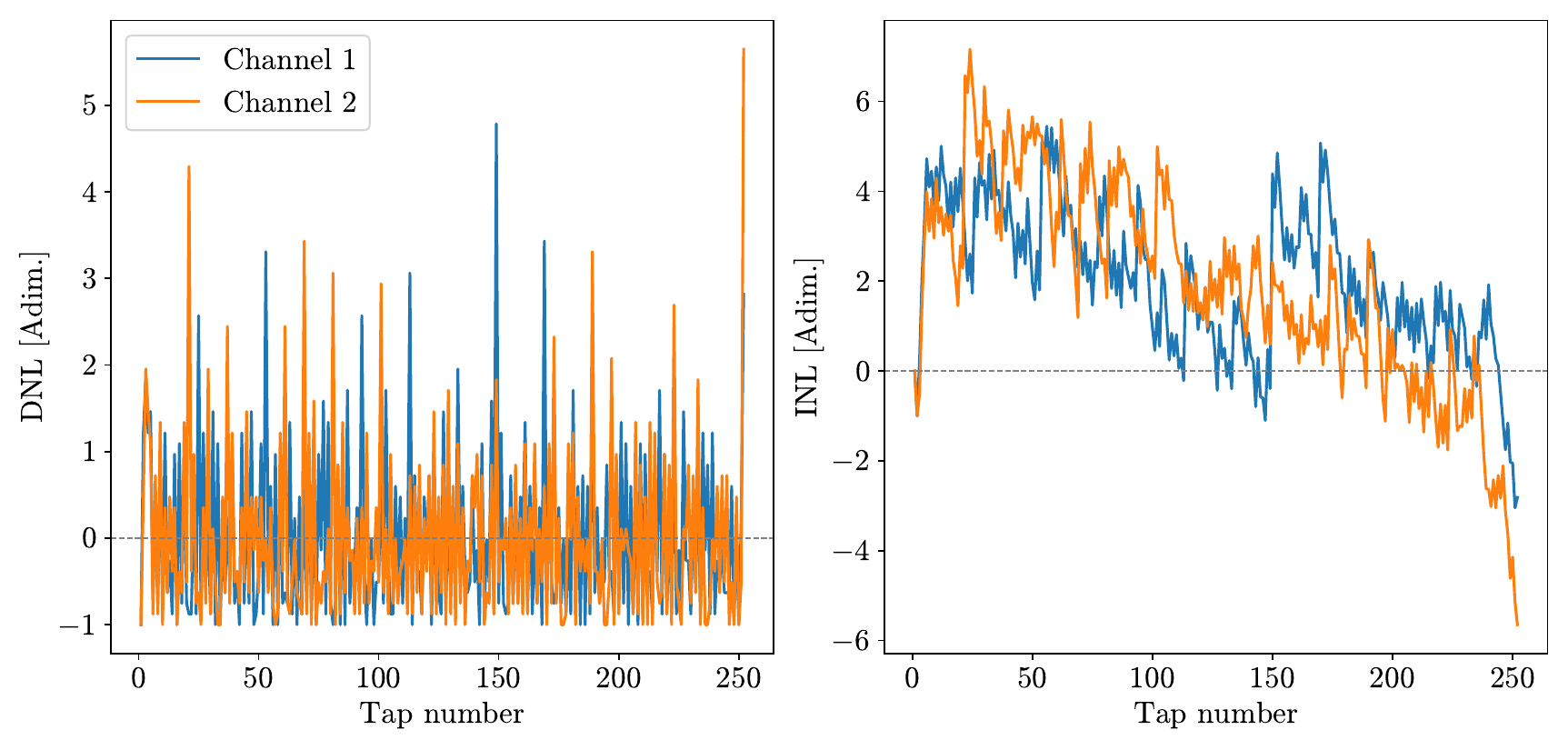}
    \caption{Differential Non-Linearity (left) and Integral Non-Linearity (right) for both channels. The DNL shows the individual tap variation with respect to the average delay $\tau_p = 19.8$~ps. A tap with DNL = +1 means that that tap has a delay of 2$\tau_p$ = 39.6~ps. The INL shows the cumulative variation in units of $\tau_p$.}
    \label{fig:non_linear}
\end{figure}

The results for the measured uncalibrated DNL and INL for both channels are shown in Figure~\ref{fig:non_linear}. The uncalibrated maximum DNL/INL measured was DNL$_{\text{max}} = +5.8$ (116~ps) and INL$_{\text{max}} = +5.4$ (108~ps) for channel 1, and DNL$_{\text{max}} = +6.6$ (132~ps) and INL$_{\text{max}} = +7.1$ (142~ps) for channel 2. The calibration allows for the correction of these effects in the other operation modes. The values obtained are consistent with uncalibrated TDLs reported in other works~\cite{nonl}. The uncalibrated performance of channel 2 is worse than channel 1, which could be due to how the TDLs are placed with respect to the TDC inputs in the design.

\begin{figure}[!h]
    \centering
    \includegraphics[width=0.7\textwidth]{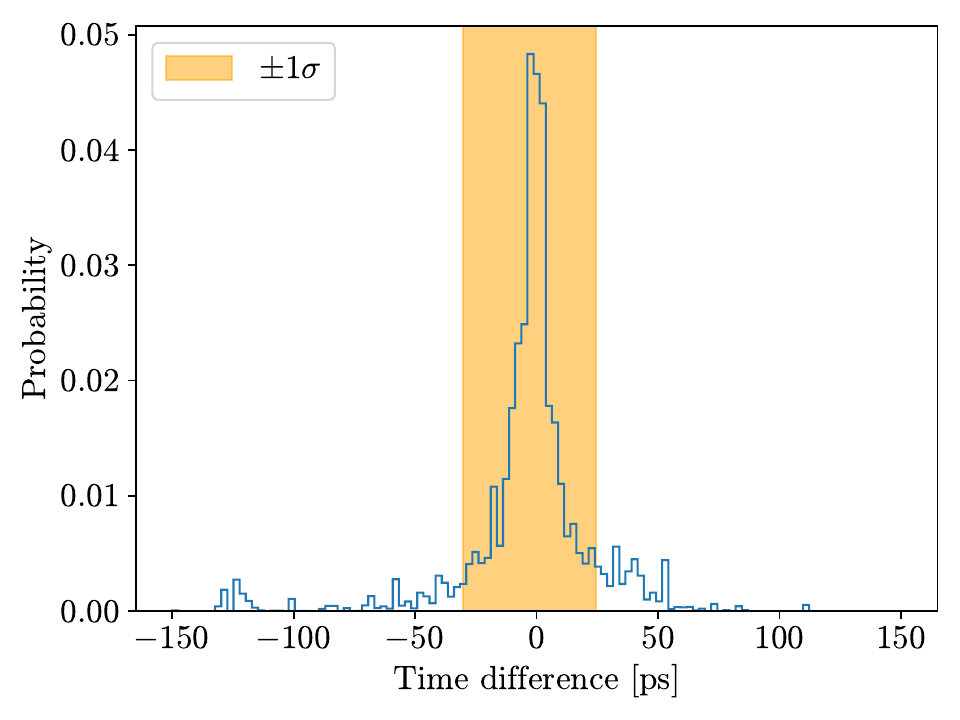}
    \caption{Time difference between channels for the same physical input. The uncertainty in this measurement is due to the FPGA design itself and sets the time resolution of the system. To get the resolution for each channel, the standard deviation of this histogram is divided by $\sqrt{2}$. The time resolution for each individual channel was then measured to be $(19 \pm 1)$~ps.}
    \label{fig:time_res}
\end{figure}

In addition, the time resolution of the system was measured and the results are shown in Figure~\ref{fig:time_res}. The uncertainty in this measurement is due to the FPGA design itself and sets the time resolution of the system. To get the resolution for each channel, the standard deviation of this histogram is divided by $\sqrt{2}$. This results in a time resolution for each individual channel of $(19 \pm 1)$~ps. The uncertainty for this measurement was obtained by bootstrapping the histogram in Figure~\ref{fig:time_res}.

\subsection{TDC and TAC measurements}

\begin{figure}[H]
    \centering
    \includegraphics[width=0.75\textwidth]{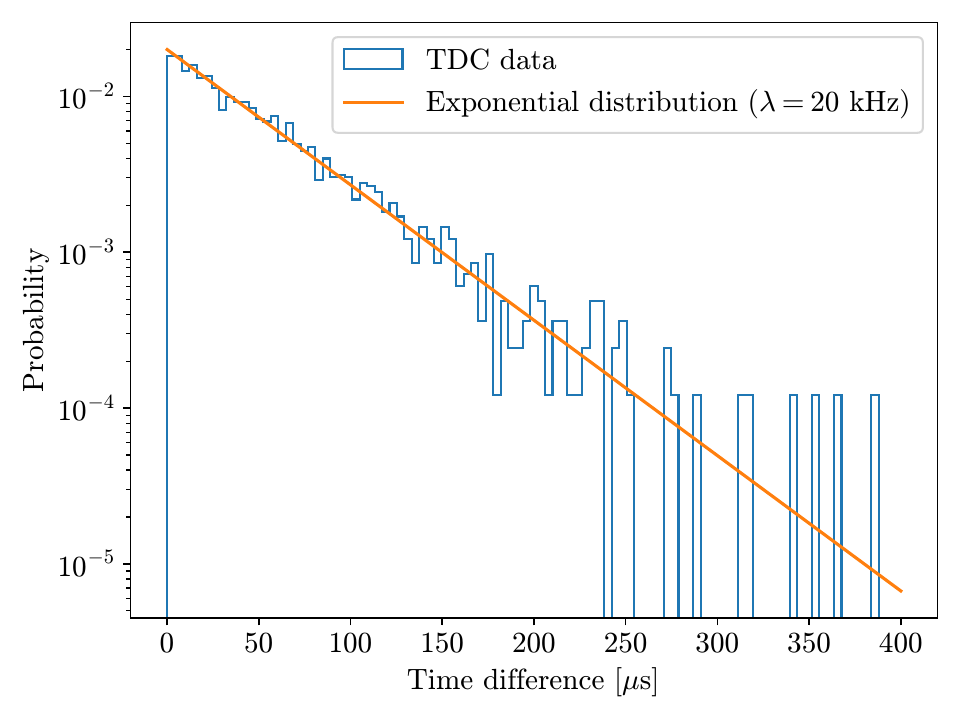}
    \caption{Time difference between subsequent events using channel 1 (after calibration), along with the expected exponential distribution. The p-value calculated by comparing measurement and model was 0.56.}
    \label{fig:exp_dist}
\end{figure}

An example of a the time difference between subsequent events in the same channel (TDC-only measurement performed using channel 1) is shown in Figure~\ref{fig:exp_dist}. It can be seen that the TDC-only measurements are compatible with the expected distribution for the time difference between Poisson-distributed events (which is an exponential distribution with $\lambda = 20$~kHz). The p-value was calculated comparing the measurement and theory and a value of 0.56 was obtained, showing compatibility between both.

\begin{figure}[H]
    \centering
    \includegraphics[width=0.85\textwidth]{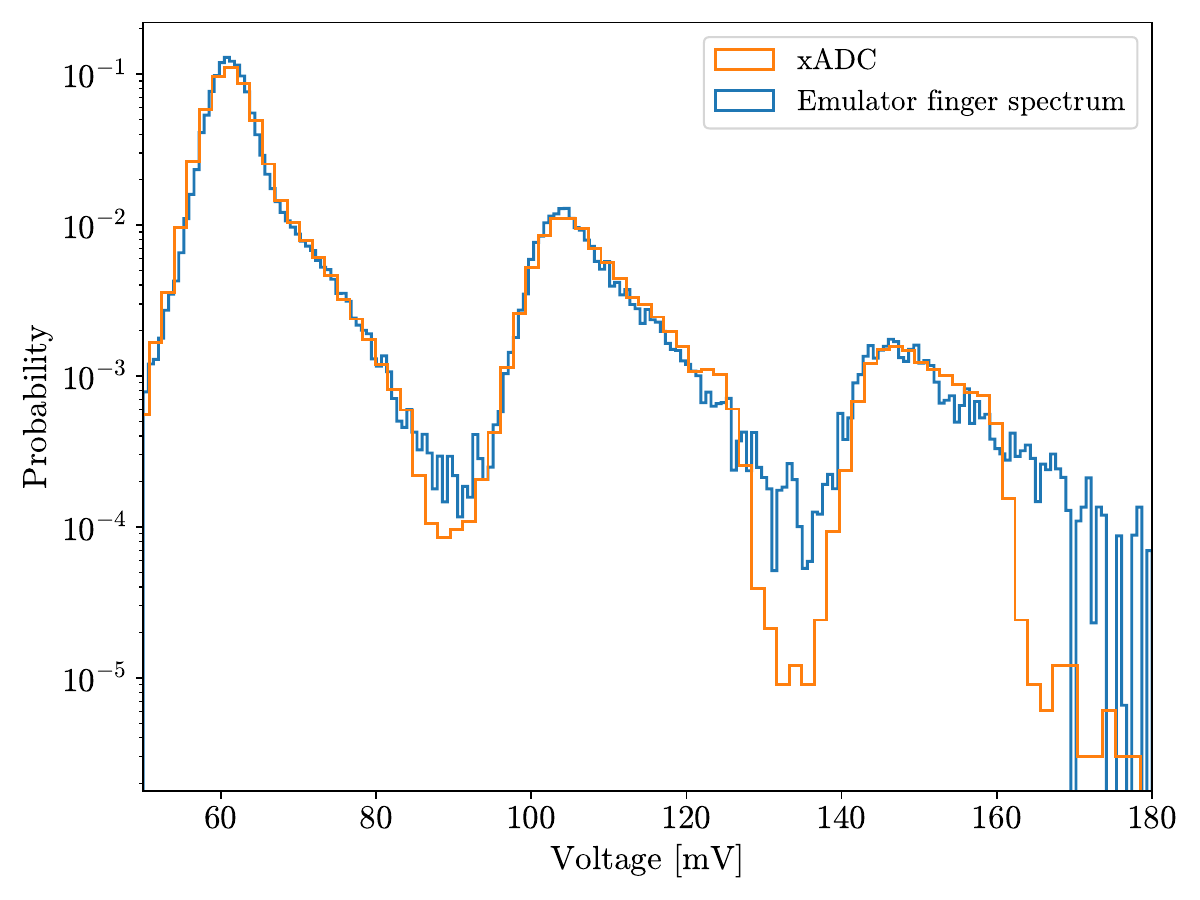}
    \caption{Amplitude measurements using the xADC, with the normalized finger spectrum used to generate the input pulses. A good agreement can be seen between these two quantities near finger spectrum peaks. However, some discrepancies exist for events with more than 2 p.e.}
    \label{fig:finger}
\end{figure}

Amplitude measurements using the xADC capabilities of the design were performed as well. These measurements and the finger spectrum used to generate the input pulses are shown in Figure~\ref{fig:finger}.

A good agreement can be seen between the amplitude measurements performed and the finger spectrum used for pulse generation. However, some discrepancies exist for events with more than 2 p.e. Due to the xADC having a 1~MSps resolution and being used in event-driven mode, the sampling instant for each analog pulse is user-defined. This makes it so pulses with higher p.e. experience time walk and the amplitude sampling is no longer performed at the maximum value, but but at an intermediate amplitude value. To circumvent this problem, and remove the fixed user-defined sampling instant, there are two possible options:

\begin{enumerate}
    \item \textbf{Peak detector circuit:} This circuit charges a capacitor during the pulse rising edge and holds its maximum value. This would allow measuring the maximum amplitude and would eliminate time-walk concerns, as the measurement can be performed when the user decides, as long as it is after the pulse rise time. Also, this would allow the measurement of much shorter pulses than the ones used in this work, as the xADC settling time is approximately 380~ns~\cite{xadc}. Without this circuit, the amplitude of SiPM pulses shorter than 380~ns can't be measured properly due to this limitation.
    \item \textbf{External ADC:} An external ADC would remove the need for a peak detector circuit and would allow for faster waveform digitalization. However, this adds additional complexity to the design, as a communication with the external ADC would need to be implemented in place of the xADC. 
\end{enumerate}

For 2~$\mu$s FWHM pulses or broader, this design allows for the correct discrimination of pulse types of different photoelectrons. However, one of the two alternatives presented above need to be considered for faster pulses due to xADC sampling limitations.

\subsection{COINCIDENCE mode measurements}

\begin{figure}[H]
    \centering
    \includegraphics[width=0.75\textwidth]{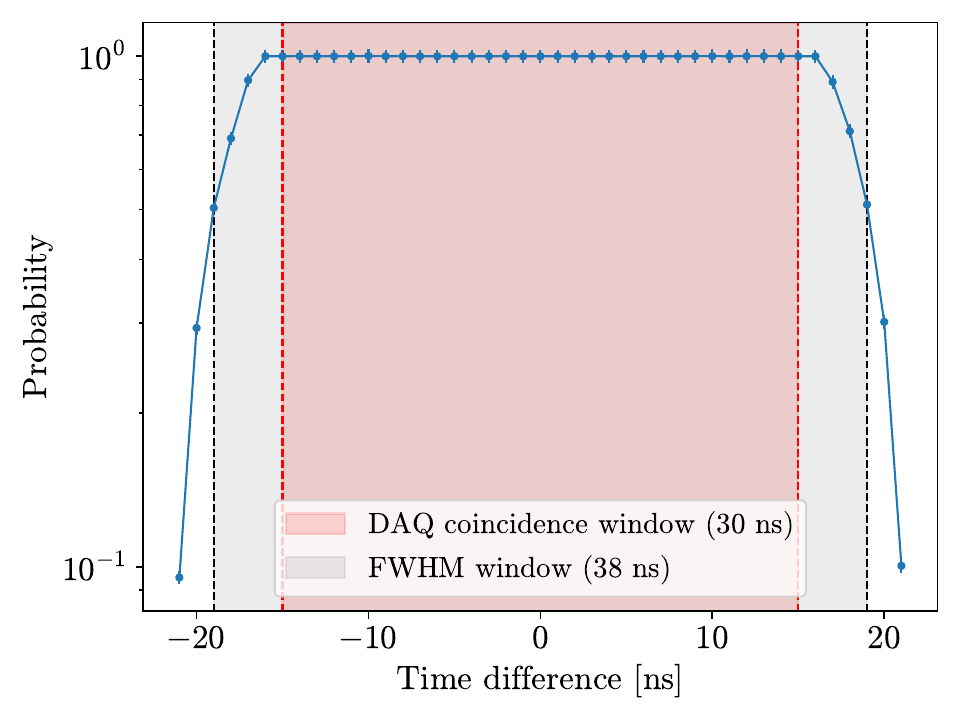}
    \caption{Detection probability for a $\pm$15~ns user-defined DAQ coincidence window. The detection probability is 1 inside the window and falls to zero outside. The physical FWHM window is 8~ns larger than the user-defined coincidence window. This effect can be mitigated by choosing a smaller coincidence window.}
    \label{fig:coinc_window}
\end{figure}

Finally, measurements were performed in coincidence and the detection probability for a $\pm$15~ns coincidence window is shown in Figure~\ref{fig:coinc_window}. The minimum coincidence window configuration with this design is 10~ns (symmetric, $\pm$5~ns) and is given by the system clock of the FPGA hardware design (200~MHz). The coincidence window can be increased in symmetric $\pm5$~ns steps. Figure~\ref{fig:coinc_window} shows that the non-coincident event rejection is not binary and there is a probability to detect an event which is outside this window. The FWHM physical window is shown to be $\pm4$~ns larger than the user-defined DAQ window. Events which have timing separations larger than this physical window are greatly suppressed. The effect of the physical window being larger than the used-defined window is present, but can be compensated by choosing a smaller coincidence window to offset this value. This measurement shows, however, that the minimum FWHM coincidence window achievable is $\pm 9$~ns.

\section{Conclusions and Outlook} \label{sec:conclusions}

In this work, a TDC design with amplitude measurement capabilities was presented and implemented in an Artix-7 XC7A35T part. For amplitude measurements, the internal xADC of the FPGA was chosen over other alternatives for simplicity and to make the design software defined. This design has two channels and can operate in timestamp-only mode or timestamp + amplitude mode. In turn, both channels can be operated independently or in coincidence.

This design was tested using an Arty A7-35T development board and a CAEN DT5810 pulse emulator for input signal generation. The INL and DNL were studied, along with the intrinsic time resolution of the TDC, which was measured to be $(19 \pm 1)$~ps. Timing measurements were performed using a 20~kHz Poisson-distributed source to showcase the timing capabilities of the TDC, and these were compared successfully with the expected exponential distribution. Amplitude measurements were also shown and compared with the expected finger spectrum used to generate input pulses. The xADC measurements agree with the expected spectrum in most of their range. However, the amplitude measurement is inaccurate for higher p.e. events. This is due to pulse time walk and the fact that the internal xADC (which has 1~MSps and 380~ns settling time) was used as a tradeoff for design simplicity. The coincidence mode was also shown to work with successful results, even though the FWHM window is larger the the user-defined DAQ window.

In a future design, an external ADC with a sampling rate of $\sim$100 MSps will be used to be able to avoid the time walk issues and to be able to use faster input pulses. In addition, a custom PCB will be manufactured for this design, and this will allow to bypass pMOD connector bandwidth limitations when using fast digital/analog signals.

\section*{Acknowledgements}

The authors acknowledge financial support from \mbox{ANPCyT PICT 2017-0984} ``Componentes Electrónicos para Aplicaciones Satelitales (CEpAS)'', \mbox{PICT-2019-2019-02993} \mbox{``LabOSat:} desarrollo de un Instrumento detector de fotones individuales para aplicaciones espaciales'' and \mbox{UNSAM-ECyT} FP-001.

\printbibliography

\end{document}